\documentclass[12pt]{iopart}
\usepackage{graphicx}
\usepackage{bm}
\usepackage{cite}
\usepackage{amssymb, dsfont}
\usepackage{booktabs}

\begin{document}

\title{Run-and-tumble particles, telegrapher's equation and absorption problems with 
partially reflecting boundaries}

\author{Luca Angelani}

\address{
Istituto dei Sistemi Complessi-Consiglio Nazionale delle Ricerche (ISC-CNR, UOS Sapienza)
and Dipartimento di Fisica, Universit\`a {\it Sapienza}, Piazzale Aldo Moro 2, 00185 Rome, Italy}

\ead{luca.angelani@phys.uniroma1.it}

\begin{abstract}
Absorption problems of 
run-and-tumble particles, 
described by the telegrapher's equation, 
are analyzed in one space dimension considering partially reflecting boundaries.
Exact expressions for the probability distribution function in the Laplace domain and
for the mean time to absorption are given, discussing some interesting limits 
(Brownian and wave limit, large volume limit) and different case studies
(semi-infinite segment, equal and symmetric boundaries, totally/partially reflecting boundaries).
\end{abstract}

\maketitle

\section{Introduction}

Determining the first-passage time is a key issue in the analysis of many stochastic processes \cite{redner}.
How long a signal takes to reach a threshold value or
a diffusive particle to reach a target, to escape
from a confined region through a small aperture, to be absorbed 
by a particular site of a boundary 
are very general questions which apply to
different phenomena in the real world, 
from chemical reactions \cite{kramers,hanggi}
and intracellular transport \cite{fpt_it}
to animal movements \cite{fpt_am} and financial time series \cite{FPT_book}.
Strictly speaking the problem is to determine the distribution of the
time a particle takes to reach for the first time a given point.
However, in many real situations the boundary is not perfectly absorbing
and the particle can reach a given site many times before reacting with it.
In such a case the relevant quantity is not the {\it first} time the particle
reaches the site, but, instead, the time the particle {\it reacts} with it.
Depending on the different contexts, such as, for example, absorption processes,
chemical reactions, escaping from confined regions, one calls this quantity
absorption time, reaction time, exit time, escape time.
In the following we adopt the term  {\it time to absorption} or
{\it absorption time},
having in mind that, in the presence of partially absorbing boundaries, 
strictly speaking this is not the first time the particle reaches the site,
but the first time the particle reacts with it (which are the same in the case
of perfectly absorbing boundary).
We are then dealing with {\it partially} reflecting (or absorbing) boundaries,
the so called radiation boundary conditions.
While such problems have been widely studied in the case of diffusive phenomena 
\cite{redner,FPT_book,ben1},
a very good approximation for many physical and chemical processes,
less attention, up to now, has been put on the non-diffusive case.
This is the case, for example, of search processes with  non-diffusive relocations 
in the presence of finite reaction rate  \cite{ben2},
or processes involving  active particles \cite{active}, 
like  self-propelled cells \cite{berg}, 
Janus particles \cite{janus,janus1,janus2} or 
molecular motors in microtubular filaments \cite{filaments}.
Active matter is becoming a relevant and growing field of research and allows us to extend
many tools of statistical mechanics to the fascinating field of biological non-equilibrium
processes \cite{cates}.
Strictly related are the processes governed by the telegrapher's equation \cite{weiss},
which in 1D corresponds exactly to the run-and-tumble motion of a self-propelled particle.
As an example, the firing of neurons can be modeled as a stochastic process with dichotomous noise,
resulting in a telegrapher-like asymmetric equation \cite{neurons}.
First-passage time problems and solutions of the telegrapher's equation 
in the presence of totally absorbing and reflecting boundaries have been analyzed in the past 
\cite{weiss84,maso1992,ley93}.
More specifically, in Ref. \cite{maso1992} the solution for the probability density of the particle displacement 
is reported for one and two absorbing boundaries; 
different combinations of reflecting/absorbing boundaries are analyzed in Ref. \cite{ley93}.
Perfectly absorbing boundaries are also considered in a recent paper analyzing first-passage
time problems of run-and-tumble particles \cite{epje2}.
Less attention, instead, has been paid to the problems with partially absorbing boundaries \cite{weiss}.
In Ref.  \cite{maso1993} a  solution of the telegrapher's equation is given,
considering the special case of one partly absorbing point. \\
The aim of the present paper is to investigate absorption problems described by 
the telegrapher's equation in the presence of partially reflecting boundaries.
Exact expressions of the distribution (in the Laplace domain) of absorption time and 
its mean value will be given for the general situation of 
two different partially reflecting/absorbing boundaries, 
recovering some previously obtained results as particular cases.

The paper is organized as follow.
In Sec. 2 the run-and-tumble model is introduced together with the appropriate
boundary conditions.
In Sec. 3 we define the time to absorption, giving a general expression
for its mean value.
In Sec. 4 and 5 two limiting cases are discussed, respectively the diffusive limit and the wave limit.
In Sec. 6 different case studies are reported, giving exact expressions
for the Laplace transformed probability distribution and the mean
time to absorption.
In Sec. 7 we discuss some features of the survival probability and absorption time distribution
(higher moments, large volume limit) 
in the simplified case of symmetric boundaries.
Conclusions are drawn in Sec. 8.

\section{Run-and-tumble model and telegrapher's equation}

We consider a run-and-tumble particle in one dimension confined in a finite 
interval $[a,b]$.
The particle moves at constant speed $v$ and  performs tumble events
at rate $\alpha$, randomly reorienting its direction of motion.
We consider here the case of instantaneous tumbling,
a simplified description of realistic run-and-tumble bacteria \cite{berg}.
The case of finite tumbling time can be treated in the framework of bimodal processes,
and analytical results have been recently obtained for free run-and-tumble particles  \cite{ange13,det14}.
Denoting with $P_{_R}(x,t)$ and $P_{_L}(x,t)$ the probability density function
(PDF), respectively, of 
right-oriented and left-oriented particles,
we can write the continuity equations as 
\cite{epje2,sch_1993,tail,tail2,mart,LA_epl}
\begin{eqnarray}
\label{eq_r}
\frac{\partial P_{_R}}{\partial t} &=& - v \frac{\partial P_{_R}}{\partial x}
- \frac{\alpha}{2} P_{_R} + \frac{\alpha}{2} P_{_L} \\  
\label{eq_l}
\frac{\partial P_{_L}}{\partial t} &=& v \frac{\partial  P_{_L}}{\partial x}
+ \frac{\alpha}{2} P_{_R} - \frac{\alpha}{2} P_{_L} 
\end{eqnarray}
By introducing the total PDF $P$ and the current $J$
\begin{eqnarray}
\label{P}
P  &=&  P_{_R}  +  P_{_L}  \\  
J  &=&  v ( P_{_R}  -  P_{_L} ) 
\label{J}
\end{eqnarray}
the equations for run-and-tumble particles can be written as
\begin{eqnarray}
\label{eq_p}
\frac{\partial P}{\partial t} &=& - \frac{\partial J}{\partial x} \\
\label{eq_j}
\frac{\partial J}{\partial t} &=& - v^2 \frac{\partial P}{\partial x} 
- \alpha J 
\end{eqnarray}
which correspond to the telegrapher's equation for $P$ (or for $J$)
\begin{equation}
\frac{\partial^2  P}{\partial t^2} +
\alpha \frac{\partial  P}{\partial t} =
v^2 \frac{\partial^2  P}{\partial x^2}
\label{tel}
\end{equation}
By using the Laplace transform in the time domain
\begin{equation}
{\tilde P}(x,s) = \int_0^\infty dt\ e^{-st} \ P(x,t)
\end{equation}
the Eqs. (\ref{eq_p},\ref{eq_j}) become
\begin{eqnarray}
\label{eq_p_s}
- P(x,0) + s {\tilde P}(x,s)  &=& - \frac{\partial {\tilde J}}{\partial x}(x,s) \\  
\label{eq_j_s}
- J(x,0) + s {\tilde J}(x,s)  &=& - v^2 \frac{\partial {\tilde P}}{\partial x}(x,s) - \alpha  {\tilde J}(x,s)
\end{eqnarray}

\subsection{Initial conditions}
We consider initial conditions
\begin{eqnarray}
P_{_R} (x,0) &=& \frac{\delta(x)}{2}\\
P_{_L} (x,0) &=&  \frac{\delta(x)}{2}
\end{eqnarray}
corresponding to equally distributed left and right oriented particles
at the origin $x=0$ (in the following we assume, without loss of generality, that $a<0<b$).
Initial conditions for $P$ and $J$ read
\begin{eqnarray}
\label{Pic}
P(x,0) &=&  \delta(x)\\
J(x,0) &=& 0
\end{eqnarray}
and Eq.s (\ref{eq_p_s},\ref{eq_j_s}) become
\begin{eqnarray}
\label{eq_p_s2}
&\frac{\partial {\tilde J}}{\partial x} =& \delta(x) - s {\tilde P} \\  
\label{eq_j_s2}
v^2 &\frac{\partial {\tilde P}}{\partial x} =& - (s+\alpha)  {\tilde J}
\end{eqnarray}
which correspond to the second order differential equation for ${\tilde P}$
\begin{equation}
v^2 \frac{\partial^2 {\tilde P}}{\partial x^2} -s (s+\alpha)  {\tilde P}
= -(s+\alpha) \delta(x)
\label{eq2P}
\end{equation}
In Sec. 3 we will extend the analysis to the case of general initial conditions.

\subsection{Boundary conditions}
We consider partially reflecting (partially absorbing) boundaries, 
with reflection coefficient $\gamma$ (absorption coefficient $\eta=1-\gamma$).
This corresponds to particles that, arriving at boundaries, can be either reflected, 
instantaneously reverting their direction of motion, or absorbed. 
It is worth noting that this does not encompass the situation in which particles
can be stuck at a confining wall due to their persistent motion, which requires a different treatment.
Boundary conditions at $a$ and $b$ can be written as \cite{maso1993,weiss}
\begin{eqnarray}
P_{_R}(a,t) &=&  \gamma_a P_{_L}(a,t)\\
P_{_L}(b,t) &=&  \gamma_b P_{_R}(b,t)
\end{eqnarray}
where we have considered the possibility that the two boundaries can be different,
with reflection coefficients $\gamma_a$ and $\gamma_b$.
In terms of $P$ and $J$ we have
\begin{eqnarray}
J(a,t) &=&  - \epsilon_a v P(a,t)\\
J(b,t) &=&    \epsilon_b v P(b,t) \label{bc_b}
\end{eqnarray}
where we have introduced the coefficient $\epsilon$
\begin{equation}
\epsilon = \frac{1-\gamma}{1+\gamma} = \frac{\eta}{2-\eta}
\label{epsilon}
\end{equation}
For $\gamma=0$ ($\eta=\epsilon=1$) the boundary is perfectly absorbing,
while for $\gamma=1$ ($\eta=\epsilon=0$) the boundary is perfectly reflecting.

In the Laplace domain  boundary conditions read:
\begin{eqnarray}
\label{bc1}
v \left. \frac{\partial {\tilde P}}{\partial x} \right|_{x\!=\!a} =  (s+\alpha) \epsilon_a {\tilde P}(a,s) \\
v \left. \frac{\partial {\tilde P}}{\partial x} \right|_{x\!=\!b} = - (s+\alpha) \epsilon_b {\tilde P}(b,s)
\label{bc2}
\end{eqnarray}

The problem is then to find the solution of Eq. (\ref{eq2P}) with boundary conditions (\ref{bc1},\ref{bc2}).
The solution is of the form 
\begin{equation}
{\tilde P}(x,s) = \left\{ 
\begin{array}{r l}
& A_1 e^{cx} + A_2 e^{-cx} \hspace{2cm} {\mbox{for $x>0$}} \\
& A_3 e^{cx} + A_4 e^{-cx} \hspace{2cm} {\mbox{for $x<0$}} \\
\end{array} \right.
\end{equation}
where
\begin{equation}
v^2 c^2 = s (s+\alpha)
\label{cdef}
\end{equation}
Coefficients $A_i$ ($i=1,\dots,4$) are determined by imposing continuity of ${\tilde P}$
and discontinuity of $\partial_x {\tilde P}$ -- from Eq. (\ref{eq2P}) -- 
at $x=0$ and boundary conditions at $x=a$ and $x=b$, Eq.s (\ref{bc1},\ref{bc2}).


\section{Time to absorption}

The survival probability, i.e. the probability that the particle has not yet been absorbed at time $t$,
is 
\begin{equation}
\mathbb{P}(t)=\int_a^b dx \ P(x,t)
\label{survP}
\end{equation}
The PDF of the absorption time 
is
\begin{equation}
\varphi (t) = - \frac{\partial \mathbb{P}}{\partial t} (t)
\label{pdf_phi}
\end{equation}
and its  Laplace transform reads
\begin{equation}
{\tilde \varphi} (s) = 1 - s \int_a^b dx \ {\tilde P}(x,s) 
= 1 - s {\tilde  \mathbb{P}}(s)
\label{pdfs_phi}
\end{equation}
By using Eq. (\ref{eq_p_s2}) it is easy to show that 
\begin{equation}
{\tilde \varphi} (s) ={\tilde J} (b,s) - {\tilde J} (a,s)
\end{equation}
or, in term of ${\tilde P}$
\begin{equation}
{\tilde \varphi} (s) =\epsilon_b v {\tilde P} (b,s) + \epsilon_a v {\tilde P} (a,s)
\end{equation}
The time to absorption distribution is then obtained from the  solution of Eq. (\ref{eq2P})
calculated at boundaries $a$ and $b$.
After some algebra one finally obtains 
\begin{equation}
{\tilde \varphi} (s) = \frac{
\epsilon_a (\epsilon_b + \frac{s}{vc}) e^{cb} -
\epsilon_a (\epsilon_b - \frac{s}{vc}) e^{-cb} -
\epsilon_b (\epsilon_a - \frac{s}{vc}) e^{ca} +
\epsilon_b (\epsilon_a + \frac{s}{vc}) e^{-ca} 
}
{
(\epsilon_a + \frac{s}{vc}) (\epsilon_b + \frac{s}{vc}) e^{c(b-a)} -
(\epsilon_a - \frac{s}{vc}) (\epsilon_b - \frac{s}{vc}) e^{-c(b-a)} 
}
\label{phis}
\end{equation}
or, rearranging the terms
\begin{equation}
{\tilde \varphi} (s) = \frac{
\epsilon_a \epsilon_b v c  (s +\alpha) (\sinh cb - \sinh ca)+
s (s+\alpha) (\epsilon_b \cosh ca + \epsilon_a \cosh cb)
}
{
[\epsilon_a \epsilon_b (s+\alpha) + s] vc \sinh c(b-a) +
s (s+\alpha) (\epsilon_a+\epsilon_b) \cosh c (b-a)
}
\label{phis2}
\end{equation}
The meaning of the different parameters of the model are summarized in Table \ref{tab1}.
It is worth noting that the quantity $c$ is a function of $s$ and depends on the parameters
$v$ and $\alpha$ through the Eq. (\ref{cdef}).

\begin{table}[t!]
\begin{center}
\caption{Parameters of the model.}
\label{tab1}
\begin{tabular}{ccl}
\toprule
Quantity & & Meaning\\
\midrule
$v$ & & particle's speed \\
$\alpha$ & & particle's tumbling rate \\
$a,b$ & & left and right position of boundaries ($a<0<b$) \\
$\epsilon_a, \epsilon_b$ & & absorption coefficients of boundaries 
[see Eq.(\ref{epsilon})] \\
\bottomrule
\end{tabular}
\end{center}
\end{table}

\noindent
The mean time to absorption $\tau$ is defined by
\begin{equation}
\tau = \int_0^\infty dt \ t \ \varphi(t)
\end{equation}
and can be obtained from the derivative of $\tilde{\varphi}$
\begin{equation}
\tau = - \left. \frac{\partial {\tilde \varphi}}{\partial s} \right|_{s\!=\!0}
\end{equation}
By using the previous expressions one finally obtains 
\begin{equation}
\tau = \frac{1}{2v^2} \ \frac{
(b-a) (2v^2 - \epsilon_a \epsilon_b a b \alpha^2) +
\alpha v [\epsilon_a a^2 +\epsilon_b b^2 -2 a b (\epsilon_a + \epsilon_b)]
}
{
v (\epsilon_a +  \epsilon_b) + \epsilon_a  \epsilon_b  \alpha (b-a) 
}
\label{tau}
\end{equation}

\noindent
The above expression has been obtained considering the initial condition $P(x,0)=\delta(x)$,
Eq. (\ref{Pic}).
It is easy to generalize it to the case of generic initial conditions.
By considering particles starting at a generic point, 
$P(x,0)=\delta(x-x_0)$, and explicitly writing the dependence on boundaries $a,b$ and initial point 
$x_0\in[a,b]$, we  can write 
\begin{equation}
\tau(a,b,x_0)=\tau(a-x_0,b-x_0,0)
\end{equation}
The average over generic initial conditions with distribution $p_{in}(x_0)$ is then obtained as
\begin{equation}
\langle \tau \rangle = \int_a^b dx_0 \ p_{in}(x_0)\ \tau(a-x_0,b-x_0,0)
\end{equation}
where the expression of $\tau$ inside the integral is that of Eq. (\ref{tau}) with
the substitution $a \to a-x_0$ and $b \to b-x_0$.
For example, considering uniform initial conditions $p_{in}(x_0)=1/(b-a)$, one obtains
\begin{equation}
\langle \tau \rangle = \frac{b-a}{12 v^2}\
\frac{12 v^2  +
\alpha (b-a) [ 4 v (\epsilon_a + \epsilon_b) + \epsilon_a \epsilon_b \alpha (b-a) ]
}
{
v (\epsilon_a +  \epsilon_b) + \epsilon_a  \epsilon_b  \alpha (b-a) 
}
\label{tau_av}
\end{equation}

\section{Brownian limit}

The Brownian (diffusion) limit is obtained by considering both 
large speed and tumble rate at constant diffusivity:
$v \to \infty$, $\alpha \to \infty$, $D=v^2/\alpha$. 
The telegraph equation reduces to the diffusion (Fokker - Planck) equation
\begin{equation}
\frac{\partial  P}{\partial t} =
D \frac{\partial^2  P}{\partial x^2}
\label{fpeq}
\end{equation}
Partially reflecting boundaries for Brownian particles are described by the so called radiation boundary condition \cite{redner}
\begin{equation}
J(b,t) = k P(b,t)
\end{equation}
where the  coefficient $k$ is in the range $[0,\infty)$, $k=0$ describing the reflecting case
and $k \to \infty$ the absorbing one.
By comparing with Eq. (\ref{bc_b}) we then have that the correct limit is obtained considering a 
velocity dependent
absorption coefficient $\eta$ (or $\epsilon$) which for large $v$ behaves as 
$\eta \simeq 2k/v$ ($\epsilon \simeq k/v$) \cite{erch}.
Possible choices are for example $\epsilon = \tanh{(k/v)}$ or $\epsilon = 1-\exp{(-k/v)}$.

We then obtain the correct expression for the PDF $\tilde{\varphi}$ in the Brownian case
from the general expression reported in the previous Section, Eq. (\ref{phis}), considering
$v,\alpha \to \infty$ with $D=v^2/\alpha$
and $\epsilon v \to k$, $c^2 \to s/D$ 
\begin{equation}
{\tilde \varphi}_{_B} (s) = \frac{
k_a (k_b + \frac{s}{c}) e^{cb} -
k_a (k_b - \frac{s}{c}) e^{-cb} -
k_b (k_a - \frac{s}{c}) e^{ca} +
k_b (k_a + \frac{s}{c}) e^{-ca} 
}
{
(k_a + \frac{s}{c}) (k_b + \frac{s}{c}) e^{c(b-a)} -
(k_a - \frac{s}{c}) (k_b - \frac{s}{c}) e^{-c(b-a)} 
}
\label{phis_bm}
\end{equation}
The mean time to absorption is obtained in the same limit from Eq. (\ref{tau})
\begin{equation}
\tau_{_B} = \frac{1}{2D} \ \frac{
(b-a)(2D^2 - k_a k_b a b)  +
D [k_a a^2 +k_b b^2 -2 a b (k_a + k_b)]
}
{
D(k_a + k_b) + k_a  k_b  (b-a) 
}
\end{equation}

\section{Wave limit}

Another interesting limit is obtained for $\alpha \to 0$, i.e. considering the absence of tumbling.
The telegraph equation in this limit reduces to the wave equation
\begin{equation}
\frac{\partial^2  P}{\partial t^2} =
v^2 \frac{\partial^2  P}{\partial x^2}
\label{weq}
\end{equation}
In such a case a particle moves at constant velocity until it hits a barrier, where, with probability $\gamma$,
reverses the direction of motion, and, with probability $1-\gamma$, is absorbed.

\noindent
One obtains in such a case:
\begin{equation}
{\tilde \varphi}_{_W} (s) = \frac{
\epsilon_a (\epsilon_b + 1) e^{sb/v} -
\epsilon_a (\epsilon_b - 1) e^{-sb/v} -
\epsilon_b (\epsilon_a - 1) e^{sa/v} +
\epsilon_b (\epsilon_a + 1) e^{-sa/v} 
}
{
(\epsilon_a + 1) (\epsilon_b + 1) e^{s(b-a)/v} -
(\epsilon_a - 1) (\epsilon_b - 1) e^{-s(b-a)/v} 
}
\label{phis_w}
\end{equation}
The mean time to absorption takes the simple form
\begin{equation}
\tau_{_W} = \frac{1}{v} \ \frac{b-a}{\epsilon_a + \epsilon_b}
\label{tauw}
\end{equation}
The above expression can be also obtained by simple considerations.
Indeed, in the absence of tumbling the particle bounces back and forth inside the box
and each time it reaches the boundary $a$ ($b$) it is absorbed with probability $\eta_a$
($\eta_b$) and reflected with probability $\gamma_a=1-\eta_a$ ($\gamma_b=1-\eta_b$).
If the particle starts its motion at the origin with right-oriented velocity, 
after a time $b/v + nL/v$ ($L=b-a$ and $n=0,1,...$)  for even $n$ it is absorbed by the boundary $b$ 
with probability $\eta_b  \gamma_b^{n/2}  \gamma_a^{n/2}$,
and for odd $n$ it is absorbed by the boundary $a$ 
with probability $\eta_a  \gamma_b^{(n+1)/2}  \gamma_a^{(n-1)/2}$.
A particles starting with left-oriented velocity, at time  
$|a|/v + nL/v$ ($a<0$) for even $n$ it  is absorbed by the boundary $a$ 
with probability $\eta_a  \gamma_a^{n/2}  \gamma_b^{n/2}$,
and for odd $n$ it is absorbed by the boundary $b$ 
with probability $\eta_b  \gamma_a^{(n+1)/2}  \gamma_b^{(n-1)/2}$.
The mean time to absorption can then be obtained as an infinite sum
\begin{equation}
\tau_{_W} = \frac{1}{2v} \sum_{n=0}^{\infty} \left[
(b+nL) \ p_n(a,b) + (|a| + nL) \ p_n(b,a) 
\right]
\end{equation}
where 
 $p_n(a,b)=\eta_b \gamma_b^{n/2} \gamma_a^{n/2}$ for even $n$ and
$p_n(a,b)=\eta_a \gamma_b^{(n+1)/2} \gamma_a^{(n-1)/2}$ for odd $n$.
The above sum gives exactly the result reported in Eq. (\ref{tauw}).

\section{Different boundaries: case studies}
We analyze here some interesting cases, giving analytic expressions 
for the time to absorption distribution in the Laplace domain 
${\tilde \varphi} (s)$
and the mean time to absorption $\tau$.

\subsection{Semi-infinite segment: $a \to -\infty$}

In the case of a semi-infinite segment $(-\infty,b]$ the expression for the 
absorption time distribution can be obtained from
Eq. (\ref{phis}) in the limit  $a \to -\infty$
\begin{equation}
{\tilde \varphi} (s) = \frac{\epsilon_b}{\epsilon_b + \frac{s}{vc}}\  e^{-cb}
\end{equation}
This situation has been treated in Ref. \cite{maso1993}, where an exact expression
for the Laplace transform of the density function $P(x,t)$ is reported and from which 
one can deduce the above expression for ${\tilde \varphi} (s)$.
It is worth noting that the time to absorption diverges
\begin{equation}
\tau \to \infty
\end{equation}
due to the open left boundary allowing particles to move far away from the absorbing wall.
\noindent
In the diffusion limit one has
\begin{equation}
{\tilde \varphi}_{_B} (s) = \frac{k_b}{k_b + \frac{s}{c}}\  e^{-cb}
\end{equation}
where $c^2=s/D$. In the case of perfect absorption ($k \to \infty$)
${\tilde \varphi}_{_B} (s) = \exp{(-b \sqrt{s/D})}$, giving rise to a
survival probability -- by performing inverse Laplace transform from Eq. (\ref{pdfs_phi}) --
$\mathbb{P}_{_B}(t) = {\mbox{erf}} (b/\sqrt{4Dt})$,
where  ${\mbox{erf}}(x)=(2/\sqrt{\pi}) \int_0^x dy \ \exp(-y^2)$ 
\cite{redner,rekra}.

\subsection{Equal boundaries: $\epsilon_a=\epsilon_b=\epsilon$}

In the case of equal boundaries, with the same reflection coefficients, we have
\begin{equation}
{\tilde \varphi} (s) = \epsilon \frac{
(\epsilon + \frac{s}{vc}) e^{cb} -
(\epsilon - \frac{s}{vc}) e^{-cb} -
(\epsilon - \frac{s}{vc}) e^{ca} +
(\epsilon + \frac{s}{vc}) e^{-ca} 
}
{
(\epsilon + \frac{s}{vc})^2  e^{c(b-a)} -
(\epsilon - \frac{s}{vc})^2 e^{-c(b-a)} 
}
\end{equation}
and
\begin{equation}
\tau = \frac{b-a}{2v\epsilon} - \frac{\alpha ab}{2v^2}
\end{equation}

\noindent
In the diffusion limit the mean time to absorption reads
\begin{equation}
\tau_{_B} = \frac{b-a}{2k} - \frac{ab}{2D}
\end{equation}

\subsection{Symmetric boundaries: $a=-b$}

If the two boundaries are equidistant from the initial particle position
(symmetric case), we have
\begin{equation}
{\tilde \varphi} (s) = \frac{
[2\epsilon_a \epsilon_b + (\epsilon_a+\epsilon_b) \frac{s}{vc}] e^{cb} -
[2\epsilon_a \epsilon_b - (\epsilon_a+\epsilon_b) \frac{s}{vc}] e^{-cb}
}
{
(\epsilon_a + \frac{s}{vc}) (\epsilon_b + \frac{s}{vc}) e^{2cb} -
(\epsilon_a - \frac{s}{vc}) (\epsilon_b - \frac{s}{vc}) e^{-2cb} 
}
\end{equation}
and
\begin{equation}
\tau = \frac{1}{2v^2}\ 
\frac{
2b (2v^2 + \epsilon_a \epsilon_b  b^2 \alpha^2) +
3 \alpha v b^2 (\epsilon_a +\epsilon_b )
}{v (\epsilon_a +  \epsilon_b)
+2 \epsilon_a  \epsilon_b  \alpha b}
\end{equation}

\noindent
In the diffusion limit the mean time to absorption is
\begin{equation}
\tau_{_B} = \frac{1}{2D}\ 
\frac{
2b (2D^2 + k_a k_b  b^2 ) +
3 D b^2 (k_a +k_b )
}{D (k_a +  k_b)
+2 k_a  k_b   b}
\end{equation}

\subsection{Equal and symmetric boundaries: $\epsilon_a=\epsilon_b=\epsilon$, $a=-b$ \label{sec_es}}

In the case of two equal and symmetric boundaries one has
\begin{equation}
{\tilde \varphi} (s) = \frac{2\epsilon}
{
(\epsilon + \frac{s}{vc}) e^{cb} +
(\epsilon - \frac{s}{vc}) e^{-cb} 
}
\end{equation}
and
\begin{equation}
\tau = \frac{b}{v\epsilon} + \frac{\alpha b^2}{2v^2} 
\end{equation}
A similar case, but considering totally absorbing boundaries ($\epsilon\!=\!1)$, 
has been analyzed in previous papers \cite{weiss84,ley93,epje2}.
The above expressions are particularly simple and 
will be taken as reference to discuss some other issues of absorption problems
in the next Section.
\noindent
In the diffusion limit the mean time to absorption reads
\begin{equation}
\tau_{_B} = \frac{b}{k} + \frac{b^2}{2D} 
\end{equation}

\subsection{One reflecting boundary: $\epsilon_a=0$}

In the case of one perfectly reflecting boundary at $x=a$ and partially reflecting at $x=b$ one has
\begin{equation}
{\tilde \varphi} (s) = \epsilon_b \frac{
e^{ca} + e^{-ca}
}
{
(\epsilon_b + \frac{s}{vc}) e^{c(b-a)} +
(\epsilon_b - \frac{s}{vc}) e^{-c(b-a)} 
}
\end{equation}
and
\begin{equation}
\tau = \frac{b-a}{v\epsilon_b} + \frac{\alpha b}{2v^2} (b-2a)
\end{equation}

\noindent 
It is worth noting that for $a=0$ previous formulae reduce to those obtained in Sec. \ref{sec_es}.
Indeed, for symmetric reasons, the problem with two equal boundaries at $-b$ and $b$ is equivalent 
to the one with a boundary at $b$ and a reflecting boundary at the origin.

\noindent
In the diffusion limit the mean time to absorption is
\begin{equation}
\tau_{_B} = \frac{b-a}{k_b} + \frac{ b}{2D} (b-2a)
\end{equation}
Considering average over uniformly distributed initial conditions 
-- see Eq. (\ref{tau_av}) -- one has
\begin{equation}
\langle \tau_{_B} \rangle = \frac{b-a}{k_b} + \frac{(b-a)^2}{3D} 
\end{equation}
in agreement with the expression reported in Table 1 of Ref. \cite{sss}.

\section{Discussion}

In this Section I discuss some properties of the absorption time probability distribution,
referring to the case of equal and symmetric boundaries, treated in the previous Section.
This allows us to obtain simple analytical expressions without losing the underlying physics.\\
The expression for the (Laplace transformed) probability distribution of the absorption time 
in the case of boundaries with the same absorption properties ($\epsilon_a=\epsilon_b=\epsilon$)
and located symmetrically with respect to the origin ($a=-b$), has been obtained in Section
 \ref{sec_es} and we rewrite it here for convenience 
\begin{equation}
\label{phitilde}
{\tilde \varphi} (s) = \frac{2\epsilon}
{
(\epsilon + \frac{s}{vc}) e^{cb} +
(\epsilon - \frac{s}{vc}) e^{-cb} }
= 
\frac{1}{\cosh cb + \frac{s}{\epsilon vc} \sinh{cb}}
\end{equation}
In the following we discuss some features of this distribution.
 
\subsection{Higher moments}
All the moments (when they exist) of the distribution can be, in principle, obtained by simple derivatives
\begin{equation}
\label{hm}
\langle t^n \rangle =  (-1)^n  \left. \frac{\partial^n {\tilde \varphi}}{\partial s^n} \right|_{s\!=\!0}
\end{equation}
The first moment is the mean time to absorption, reported in the previous Section
\begin{equation}
\label{mtta}
\tau = \langle t \rangle  = \frac{b}{v\epsilon} + \frac{\alpha b^2}{2v^2} 
\end{equation}
It is worth noting that the first term dominates when $b/\ell \ll 2/\epsilon$,
where $\ell=v/\alpha$ is the particle's mean-free path.
In other words, for small box length (with respect to $\ell$) or also 
for moderately large box length in the case of weakly absorbing boundaries ($\epsilon \ll 1$), 
the mean time to absorption is dominated by the wave-like term.
For large box length, instead, more precisely for $b/\ell \gg 2/\epsilon$, 
the second diffusive-like term dominates in Eq. (\ref{mtta}).
A more in-depth discussion of the large volume regime will be found at the end of this Section.\\
We report here also the expressions for the second moment
\begin{equation}
\label{sm}
\langle t^2 \rangle = \frac{5}{3}\tau^2 + \frac{b^2}{v^2} \left ( \frac{1}{3\epsilon^2} -1 \right) 
\end{equation}
and the third moment
\begin{equation}
\label{tm}
\langle t^3 \rangle = \frac{61}{15}\tau^3 +  \frac{\alpha b^4 }{2v^4} \left ( \frac{9}{5\epsilon^2} -5 \right) 
+\frac{b^3}{\epsilon v^3} \left ( \frac{29}{15\epsilon^2} -5 \right) 
\end{equation}
Generic expressions for higher moments will be given below in the large volume limit.

\subsection{Survival probability}
An interesting property of the survival probability $\mathbb{P}(t)$ can be inferred from the expression 
in Eq. (\ref{phitilde}), via the relation
\begin{equation}
{\tilde  \mathbb{P}}(s) = \frac{1- {\tilde \varphi} (s)}{s}
\end{equation}
We first observe that, due to the finite speed of the particle, the quantity $\mathbb{P}(t)$ 
will be $1$ up to the time $b/v$, i.e. the minimum time the particle takes to cover the distance $b$ and 
then reach the boundary. 
We can then write $\mathbb{P}(t) = 1 - \theta(t-b/v) \ Q(t-b/v)$, having introduced the quantity $Q(t)$.
Using the property 
\begin{equation}
Q (0)= \lim_{s \to \infty} s\  {\tilde Q}(s) 
\end{equation}
we can obtain the value of the survival probability at time $t=b/v$
\begin{equation}
\mathbb{P}(t\!=\!b/v) = 1- \frac{2\epsilon}{\epsilon+1}\ \exp{(-\frac{b\alpha}{2v})} 
\label{P0}
\end{equation}
The survival probability $\mathbb{P}(t)$ has then a discontinuity at $t=b/v$, jumping from $1$ to 
a lower finite value \cite{maso1992}. The previous expression has a simple physical interpretation:
the quantity $\exp{(-b\alpha / 2v)}$ is the probability that a particle has not changed 
its direction of motion up to time $t=b/v$ (thus ensuring that it reaches the boundary at time $t=b/v$)
and $\eta = 2\epsilon / (\epsilon +1)$ is the absorption probability.
Hence the second term in the right-hand side of Eq. (\ref{P0}) is the probability that 
a particle is absorbed exactly at time $t=b/v$.

\subsection{Large volume limit}

We now analyze the behavior of the probability distribution $\varphi$
in the large volume limit, $b \to \infty$.
In this limit the mean time to absorption Eq. (\ref{mtta}) diverges as
\begin{equation}
\tau \simeq \frac{\alpha b^2}{2v^2} 
\end{equation}
By defining the rescaled time ${\hat t} = t/\tau$, it is possible to show that, in the large volume limit,
the probability distribution $\psi({\hat t})$ has a universal simple form in the Laplace domain \cite{benic}
\begin{equation}
{\tilde \psi} (s) = \frac{1}{\cosh{\sqrt{2s}}} =  \textrm{sech}{\sqrt{2s}}
\end{equation}
By expanding in power of $s$ we have
\begin{equation}
{\tilde \psi} (s) = \sum_{n=0}^{\infty} \ \frac{2^nE_{2n}}{(2n)!}  \ s^n
\end{equation}
where $E_n$ are the Euler numbers.
The generic moments of the distribution are then easily obtained from Eq. (\ref{hm}) 
\begin{equation}
\langle {\hat t}^n \rangle =  (-1)^n  \frac{2^n n!}{(2n)!} \ E_{2n}
\end{equation}
For example, the first three moments are $1$, $5/3$ and $61/15$, which,
due to the relation $\langle t^n \rangle = \tau^n \ \langle {\hat t}^n \rangle$,
correspond to 
\begin{eqnarray}
 \langle t \rangle &=&   \frac{\alpha b^2}{2v^2} \\ 
 \langle t^2 \rangle &=&   \frac{5}{3} \ \tau^2 \\  
 \langle t^3 \rangle &=&   \frac{61}{15} \ \tau^3 
\end{eqnarray}
in agreement with the large volume limit of Eq.s (\ref{mtta},\ref{sm},\ref{tm}).


\section{Conclusions}

Absorption problems in the presence of partially reflecting boundaries
 have been analyzed for active particles animated by run-and-tumble 
dynamics in one space dimension (telegrapher's equation).
General expressions are obtained for the absorption time distribution in the Laplace domain
and for the mean time to absorption. Brownian and wave limits are retrieved, respectively,  
for $v,\alpha \to \infty$ and $\alpha \to 0$.
Some interesting case studies are  analyzed: the case of semi-infinite segment,
the case of equal boundaries with the same reflection coefficients, the spatially symmetric case and the case
of coexistence of totally and partially reflecting boundaries.
Large volume limit is also discussed, giving  explicit expressions for generic moments of the probability distribution
of absorption time.
Reported results can be applied to all the cases in which the system dynamics is described by a
telegraph-like equation.



\end{document}